\documentclass[prl,showpacs,superscriptaddress,twocolumn]{revtex4}

\usepackage{amsmath}
\usepackage{amsfonts}
\usepackage{amssymb}
\usepackage{graphicx}

\begin{document}

\title{Non-equilibrium Casimir-like Forces in Liquid Mixtures}

\author{T. R. Kirkpatrick}
\affiliation{Institute for Physical Science and Technology, University of Maryland, College Park, Maryland 20742, USA}
\affiliation{Department of Physics, University of Maryland, College Park, Maryland 20742, USA}

\author{J. M. Ortiz de Z\'arate}
\affiliation{Departamento de F\'{\i}sica Aplicada I, Facultad de F\'{\i}sica, Universidad Complutense, 28040 Madrid, Spain}

\author{J. V. Sengers}
\email{sengers@umd.edu}
\affiliation{Institute for Physical Science and Technology, University of Maryland, College Park, Maryland 20742, USA}

\date{\today}

\begin{abstract}
In this Letter we consider a liquid mixture confined between two thermally conducting walls subjected to a stationary temperature gradient.  While in a one-component liquid non-equilibrium fluctuation forces appear inside the liquid layer only, non-equilibrium fluctuations in a mixture induce a Casimir-like force on the walls. The physical reason is that the temperature gradient induces large concentration fluctuations through the Soret effect. Unlike temperature fluctuations, non-equilibrium concentration fluctuations are also present near a perfectly thermally conducting wall. The magnitude of the fluctuation-induced Casimir force is proportional to the square of the Soret coefficient and is related to the concentration dependence of the heat and volume of mixing.
\end{abstract}

\pacs{05.20Jj, 65.40.De, 05.70.Ln}

\maketitle

When large and long-range fluctuations are present, they will induce forces in confined fluids~\cite{REF1}. These are commonly referred to as Casimir-like forces in analogy to forces induced by vacuum fluctuations between two conducting plates~\cite{REF2,REF3}. A well-known example is the Casimir force induced by critical fluctuations in fluids~\cite{REF4,REF5,REF6,REF7}. Apart from critical systems, long-range correlations also exist in equilibrium systems with Goldstone modes~\cite{REF1} and in many non-equilibrium systems, where even longer-range correlations can exist~\cite{REF8,REF9,REF10,REF11}.

In this Letter we consider a liquid mixture in a non-equilibrium steady state (NESS) between two parallel thermally conducting plates subjected to a uniform temperature gradient $\nabla{T}$. In a liquid mixture a temperature gradient induces large concentration fluctuations through the Soret effect~\cite{REF12,REF13}. These non-equilibrium concentration fluctuations vary with the 4$^\text{th}$ power of the inverse of the wave number $k$ of the fluctuations, just as the non-equilibrium temperature fluctuations in a one-component fluid~\cite{REF8,REF14}. However, there is a principal difference between the Casimir pressures induced by non-equilibrium concentration fluctuation and those induced by non-equilibrium temperature fluctuations. In thin fluid layers, fluctuations not only may induce a force on the walls, but also may introduce an effective potential inside the fluid layer causing a modification of the density or composition profile~\cite{REF33}. While in a one-component fluid non-equilibrium fluctuations only induce the latter phenomenon yielding a re-arrangement of the density profile~\cite{REF15}, the purpose of the present letter is to demonstrate that non-equilibrium concentration fluctuations induce an actual Casimir pressure on the walls.

It is well known that in considering the dynamics of fluctuations around thermal equilibrium, nonlinear terms in the hydrodynamic equations serve to renormalize various terms in the linearized hydrodynamic equations~\cite{REF16,REF17,REF18,REF19,REF20,REF21,REF22,REF23}. Here we show that in a NESS the nonlinear terms cause a most important renormalization of the non-equilibrium (NE) pressure or normal stresses in a binary fluid. To determine the non-equilibrium induced pressure in a liquid mixture, we need to consider the pressure $p$ as a function of the fluctuating conserved quantities, which are the fluctuating energy density $e+\delta{e}$, the fluctuating mass densities $\rho_1+\delta\rho_1$, and $\rho_2+\delta\rho_2$ of components 1 (solute) and 2 (solvent). As in the case of a one-component fluid, we can neglect the fast propagating sound modes and, hence, the linear fluctuation contribution to the pressure~\cite{REF15}. Applying a Taylor expansion to the pressure then yields a contribution quadratic in terms of  $\delta{e}$,  $\delta\rho_1$, and  $\delta\rho_2$:
\begin{multline}\label{E01}
p(e+\delta{e},\rho_1+\delta\rho_1,\rho_2+\delta\rho_2)-p(e,\rho_1,\rho_2)\\=\frac{1}{2}\sum_{i,j=1}^3 \frac{\partial^2p}{\partial{a}_i\partial{a}_j}\delta{a_i}\delta{a_j}
\end{multline}
with ${\mathbf{a}}=(e,\rho_1,\rho_2)$. In a liquid mixture there are two diffusion modes that are linear combinations of heat diffusion and mass diffusion~\cite{REF12,REF24}. An important parameter for dealing with fluctuations in liquid mixtures is the Lewis number, which is the ratio of thermal diffusivity $D_T$ and mutual mass diffusivity $D$: $Le=D_T/D$. In liquid mixtures this Lewis number is commonly larger than unity. Hence, in dealing with fluctuations in liquid mixtures one often adopts a large-Lewis-number approximation~\cite{REF10}. For large values of the Lewis number, these diffusion modes decouple into a pure temperature fluctuation mode with a decay time proportional to $D_T^{-1}$ and a concentration fluctuation mode with a decay time proportional to $D^{-1}$~\cite{REF25}. Hence, to get the slowest fluctuation mode contribution for $Le\gg1$, we not only may neglect linear pressure fluctuations, but also linear temperature fluctuations.

For the concentration variable we adopt the mass fraction of the solute $w=\rho_1/\rho$ with $\rho=\rho_1+\rho_2$ being the mass density of the mixture. At constant $p$ and $T$, $\delta e,\;\delta\rho_1\;\delta\rho _2$ are related to the concentration fluctuations $\delta{w}$ by $\delta e = \left( {\partial e/\partial w} \right)_{p,T} \delta w$, $\delta\rho_1 = \left( {\partial\rho_1/\partial w}\right)_{p,T} \delta w$,  $\delta\rho_2 = \left( {\partial\rho_2/\partial w}\right)_{p,T} \delta w$.  We then obtain from Eq.~\eqref{E01} for the average NE contribution $p_{{\rm{NE}}}^w \left( \bf{r }\right)$ at a position $\mathbf{r} = \left\{ {x,y,z} \right\}$ to the equilibrium pressure $p$ in terms of $e,\rho,w$:
\begin{align}
p_{{\rm{NE}}}^w(\bf{r}) =  -& \frac{1}{2}{\left({\frac{{\partial p}}{{\partial e}}}\right)}_{\mkern-7mu\rho,w} \left[{\left({\frac{{\partial ^2 e}}{{\partial w^2 }}}\right)}_{\mkern-7mu{p,T}}\hspace*{-10pt} - {\left({\frac{{\partial{e}}}{{\partial \rho }}} \right)}_{\mkern-7mu{p,w}} {\left( {\frac{{\partial ^2 \rho }}{{\partial w^2 }}} \right)}_{\mkern-7mu{p,T}} \right.\notag\\
&-\left. \frac{2}{\rho }{\left( {\frac{{\partial e}}{{\partial w}}} \right)}_{\mkern-7mu{p,\rho}} {\left( {\frac{{\partial \rho }}{{\partial w}}} \right)}_{\mkern-7mu{p,T}}\right]
{\left\langle {\left( {\delta w\left( \bf{r} \right)} \right)^2 } \right\rangle}_{\mkern-4mu\rm{NE}},\label{E02}
\end{align}				
where the superscript $w$ indicates that $p_{{\rm{NE}}}^w \left( \bf{r }\right)$ is a Casimir pressure induced by concentration fluctuations. We note that only the NE concentration fluctuations $\langle({\delta w({\mathbf{r}}))^2}\rangle_{{\rm{NE}}}$ cause a renormalization of the pressure, since the equilibrium concentration fluctuations are already incorporated in the unrenormalized pressure. Just as for the case of a one-component fluid~\cite{REF15}, the NE pressure can be obtained from an explicit mode-coupling theory generalized to NESS, which justifies the approach adopted above.

Relevant thermodynamic relations, associated with the hydrodynamic modes in a mixture, can be found in an article of Wood~\cite{REF24}. Noting that the thermodynamic field conjugate to the mass fraction $w$ is the difference between the specific chemical potentials of the solute and the solvent, $\mu=\mu_1-\mu_2$, we can transform Eq.~\eqref{E02} into
\begin{align}\label{E03}
p_{{\rm{NE}}}^w(\bf{r})=&  - \frac{{\rho \left( {\gamma-1} \right)}}{{2\alpha T}} \left[\chi ^{ - 1} - T {\left({\frac{{\partial \chi ^{ - 1} }}{{\partial T}}} \right)}_{\mkern-7mu{p,w}}\right. \\
&- \left.\frac{{\rho c_{p,w} }}{\alpha } {\left( {\frac{{\partial \chi ^{ - 1} }}{{\partial p}}}\right)}_{\mkern-7mu{T,w}}
\right] \left\langle\left(\delta{w(\bf{r})}\right)^2\right\rangle_{\mkern-4mu\text{NE}},\notag
\end{align}
where $\alpha  =  - \rho ^{ - 1} \left( {\partial \rho /\partial T} \right)_{p,w}$ is the thermal expansion coefficient, $c_{p,w}$ the isobaric specific heat capacity, $\gamma=c_{V,w}/c_{p,w}$ the ratio of the isochoric and isobaric heat capacities, and $\chi  = \left( {\partial w/\partial \mu } \right)_{p,T}$ an osmotic susceptibility. This osmotic susceptibility can be related to the molar excess Gibbs energy~\cite{REF26} and, hence, its temperature and pressure derivatives to the excess molar enthalpy $H^{\rm{E}}$ and the excess molar volume $V^{\rm{E}}$, so that
\begin{align}\label{E04}
p_{{\rm{NE}}}^w(\mathbf{r}) &=  - \frac{{\rho \left( {\gamma  - 1} \right)}}{{2\alpha T}}\frac{{M^3 }}{{M_1^2 M_2^2 }} \\
&\times \left[ {{\left( {\frac{{\partial ^2 H^{\rm{E}} }}{{\partial x_1^2 }}}\right)}_{\mkern-7mu{p,T}}\hspace*{-10pt} - \frac{{\rho c_{p,w} }}{\alpha } {\left( {\frac{{\partial ^2 V^{\rm{E}} }}{{\partial x_1^2 }}}\right)}_{\mkern-7mu{p,T}} } \right] {\left\langle {\left(\delta{w(\mathbf{r})}\right)^2 } \right\rangle}_{\mkern-4mu{\rm{NE}}},\notag
\end{align}
where $M_1$ and $M_2$ are the molar weights of the solute and solvent, respectively, $M=M_1x_1+(1-x_1)M_2$ the molar weight of the mixture, and $x_1$ the mole fraction of the solute.

The intensity of the NE concentration fluctuations $\langle({\delta w(\mathbf{r}))^2}\rangle_{{\rm{NE}}}$ can be obtained by solving appropriate fluctuating hydrodynamics equations~\cite{REF10,REF27,REF28}. We consider a liquid mixture subjected to a stationary temperature gradient $\nabla{T}$ confined between two horizontal thermally conducting plates located at $z=\pm L/2$ in the coordinate direction perpendicular to the plates. Such a temperature gradient induces a stationary concentration gradient $\nabla w =  - w\left( {1 - w} \right)S_T \nabla T$, where $S_T$ is the Soret coefficient~\cite{REF12,REF13}. A procedure for solving the fluctuating hydrodynamic equations to obtain the intensity of the NE concentration fluctuations has been developed by two of us, but with artificial boundary conditions for the fluctuations at the walls adopted for mathematical convenience~\cite{REF28}. It turns out that exactly the same procedure can be used to obtain the solution for the intensity of the NE concentration fluctuations which satisfies physically realistic boundary conditions, namely, a rigid-boundary condition for the wall-normal velocity fluctuations, $\delta v_z  = d\delta v_z /dz = 0$~\cite{REF29}, and the condition of no mass flux through the boundaries, $d\delta w/dz = 0$, at $z=\pm L/2$, for $Le\gg1$. The solution $\langle({\delta w(\mathbf{r}))^2}\rangle_{{\rm{NE}}}$ for the intensity of the NE concentration fluctuations and, hence, for the NE Casimir pressure $p_{{\rm{NE}}}^w \left( z \right)$, only depends on the height $z$ in the liquid layer. While the solution for arbitrary values of $z$ is rather complicated, the important new result is that we have obtained a simple expression for the concentration fluctuations at the walls:
\begin{equation}\label{E05}
\left\langle {\left( {\delta w\left( { z=\pm \tfrac{L}{2}} \right)} \right)^2 } \right\rangle _{\mkern-4mu{\rm{NE}}}\hspace*{-6pt}
= \frac{{k_{\rm{B}} T}}{{\rho \nu D}}F_0 w^2 \left( {1 - w} \right)^2 S_T^2 L\left( {\nabla T} \right)^2
\end{equation}
with
\begin{equation}\label{E06}
\begin{split}
F_0  &= \frac{1}{{2\pi }}\int_0^\infty  q \left[ {\frac{1}{{q^4 }} + \frac{{4\left( {1 - \cosh \;q} \right)}}{{q^2 \left( {q + \sinh \;q} \right)}}} \right]dq
\\& \simeq 3.11 \times 10^{ - 3},
\end{split}
\end{equation}
where $k_\text{B}$ is Boltzmann's constant and $\nu$ is the kinematic viscosity. In Eq.~\eqref{E06} $q=k_\parallel L$, where $k_\parallel$ is the magnitude of the component of the wave vector $\mathbf{k}$ of the fluctuations parallel to the plates~\cite{REF28}. A derivation of the intensity of the NE concentration fluctuations, given by Eqs.~\eqref{E05} and~\eqref{E06}, can be found in~\cite{miarxiv}. Substitution of Eq.~\eqref{E05} into Eq.~\eqref{E04} yields for the NE Casimir pressure $p_{{\rm{NE}}}^w \left( {z =  \pm L/2} \right)$ exerted on the walls:
\begin{equation}\label{E07}
\begin{split}
p_{{\rm{NE}}}^w \left(z={\pm\tfrac{L}{2}} \right)&=   - \frac{{k_{\rm{B}} T^2 \left( {\gamma  - 1} \right)}}{{2\alpha \nu D}}\frac{{M^3 }}{{M_1^2 M_2^2 }}\\
&\times\left[ {{\left( {\frac{{\partial ^2 H^{\rm{E}} }}{{\partial x_1^2 }}}\right)}_{\mkern-7mu{p,T}}\hspace*{-10pt} - \frac{{\rho c_{p,w} }}{\alpha} {\left( {\frac{{\partial ^2 V^{\rm{E}} }}{{\partial x_1^2 }}}\right)}_{\mkern-7mu{p,T}} } \right]  \\
&\times F_0 w^2 \left( {1 - w} \right)^2 S_T^2 L\left( {\frac{{\nabla T}}{T}} \right)^2.
\end{split}
\end{equation}
It has been verified experimentally that approximating the thermodynamic and transport properties in Eq.~\eqref{E05} and, hence in Eq.~\eqref{E07}, by their average values in the center of the liquid layer reproduces the intensity of the non-equilibrium fluctuations to within 1\% at temperature differences up to $\Delta{T}=40$~K between two plates~\cite{REF26,REF39}. We note that for a given temperature gradient $\nabla{T}$, the Casimir pressure exerted on the walls \emph{increases} with the distance $L$ between the plates, indicating that we are dealing with a giant, \emph{i.e.}, surprisingly large, Casimir effect~\cite{REF30}. The physical reason is that the NE correlations diverge as $k^{-4}$, which means that in real space the correlations scale with the system size $L$. While there exists an extensive literature on long-range correlation in NESS, we emphasize that only NE temperature and NE concentration fluctuations cause such a dramatic effect~\cite{REF8,REF10}. Physically, it may be more practical to study the NE Casimir pressure as a function of the distance $L$  for a given temperature difference $\Delta{T}$ between the plates so that $\nabla{T}=\Delta{T}/L$. It then follows from Eq.~\eqref{E07} that $p_\text{NE}^w$ will be proportional to $L^{ - 1} \left( {\Delta T/T}\right)^2$. In principle $p_{{\rm{NE}}}^w$ is also affected by gravity. However, it is expected that this effect will be minor except when the mixture is close to a hydrodynamic instability~\cite{REF15}.

It is interesting to compare the Casimir pressure induced by NE concentration fluctuations with a Casimir pressure induced by NE temperature fluctuations in a one-component fluid~\cite{REF15}. In analogy to Eq.~\eqref{E04}, that result can be written as,
\begin{equation}\label{E08}
\begin{split}
p_{{\rm{NE}}}^T(z)& =  - \frac{{\rho \left( {\gamma  - 1} \right)}}{{2\alpha T}} \left[{\left({\frac{{\partial ^2 h}}{{\partial T^2 }}}\right)}_{\mkern-7mu{p}}-\frac{{\rho c_p }}{\alpha}
{\left(\frac{{\partial ^2 v}}{{\partial T^2 }}\right)}_{\mkern-7mu{p}}\right] \\ &\times\left\langle {\left( {\delta T(\mathbf{r})} \right)^2 } \right\rangle_{\mkern-4mu{\rm{NE}}},
\end{split}
\end{equation}
where $h$ is the specific enthalpy and $v$ the specific volume. While Eqs.~\eqref{E04} and~\eqref{E08} look very similar, there is a fundamental difference between the two. In contrast to Eq.~\eqref{E07}, Eq.~\eqref{E08} implies
\begin{equation}
p_{{\rm{NE}}}^T \left(z= { \pm \frac{L}{2}} \right) = 0,
\end{equation}
since at a thermally conducting wall $\delta{T}=0$. Hence, the Casimir pressure $p_{{\rm{NE}}}^T$ induced by NE temperature fluctuations only appears in the inside of the fluid layer, causing a re-arrangement of the density profile~\cite{REF15}, but it does not exert a pressure on the thermally conducting walls. Hence, we only quoted finite values for $\left\langle {p_{{\rm{NE}}}^T } \right\rangle _z$ averaged over the height of the fluid layer in previous publications~\cite{REF15,REF30}. We note that in the dilute-gas limit $p_{{\rm{NE}}}^T$ vanishes for any value of $z$.

\begin{figure}[b]
\begin{center}
\includegraphics[width=1\columnwidth]{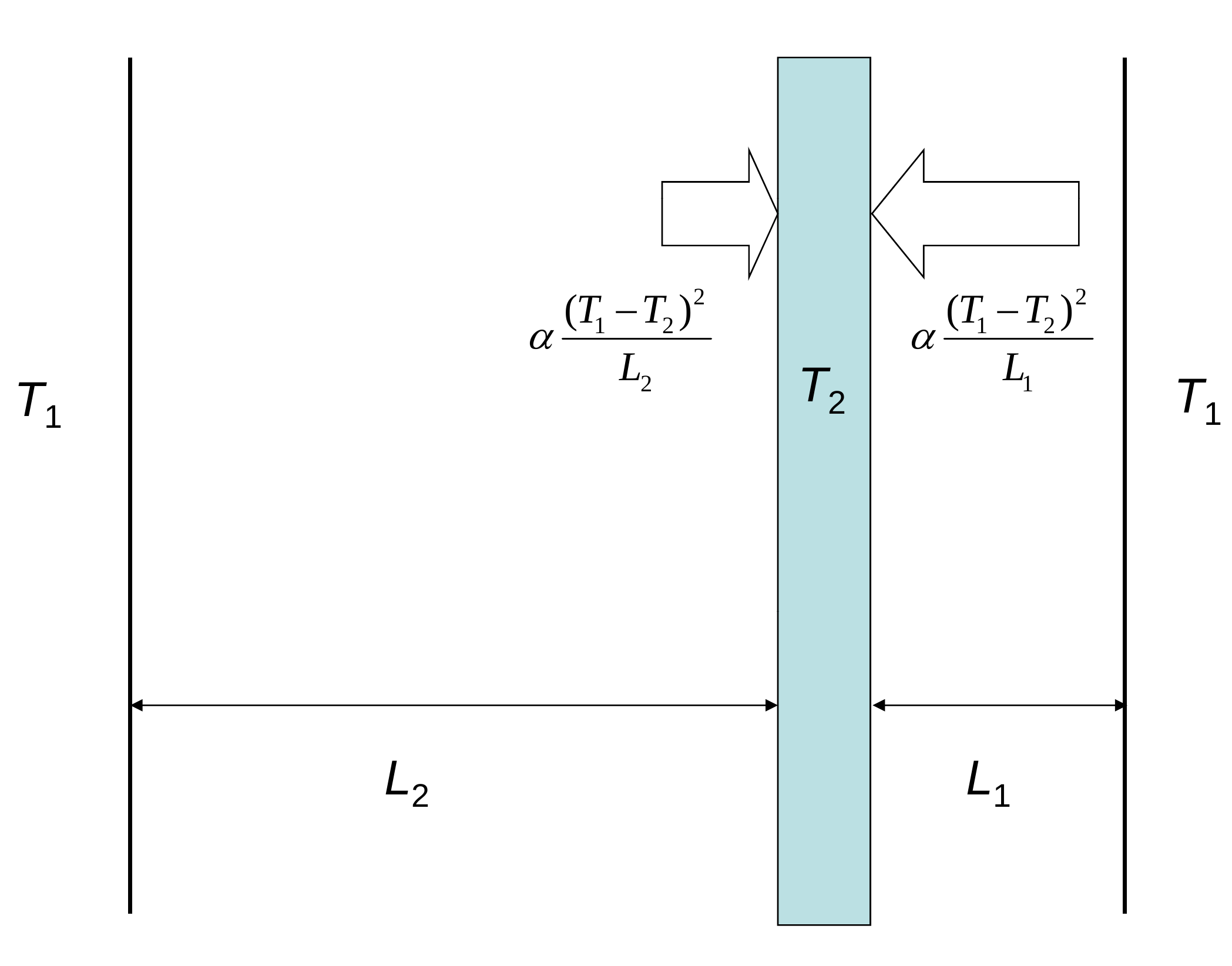}
\end{center}
\caption{Schematic illustration of Casimir pressures  $p_\text{NE}^w> 0$, induced by NE concentration fluctuations on a plate with temperature $T_2$ located in a liquid mixture between walls with temperature $T_1$. For  $p_\text{NE}^w<0$, the plate would be pulled to the closest wall.}
\label{F1}
\end{figure}

To understand the implications of Eq.~\eqref{E07}, we may envision a configuration where a (thin) plate with temperature $T_2$ is located in a liquid mixture between two walls, both with a temperature $T_1$, as schematically illustrated in Fig.~\ref{F1}.  When $p_{{\rm{NE}}}^w> 0$, the liquid mixture will exert NE Casimir pressures on the two sides of the inner plate proportional to $\left( {\Delta T} \right)^2 /L_1$ and $\left( {\Delta T} \right)^2 /L_2$. When $L_1\neq L_2$, the plate will experience a net force causing it to move to the center of the liquid mixture layer. Hence, the force needed to move this plate off center, would be a measure of the Casimir force induced by the NE concentration fluctuations. In practice it may be difficult to maintain plates at a close distance parallel to each other~\cite{REF3,REF31,REF32} and one may want to replace the plate by a particle. While a geometrical analysis of such a configuration becomes more complicated~\cite{REF5,REF33}, the physical principle remains the same.

\begin{table}
\caption{Estimated Casimir pressures}
\begin{tabular*}{\columnwidth}{l@{\extracolsep{\fill}}lcc}
\toprule
&&$L=10^{-6}$~m&$L=10^{-4}$~m \\
\colrule
$p_\text{em}$\footnote{Refs.~\cite{REF2,REF15}}&&  $-1\times10^{-3}$~Pa  &$-1\times10^{-11}$~Pa\\
$p_\text{c}$\footnote{Ref.~~\cite{REF15}}&& $-6\times10^{-4}$~Pa &$-6\times10^{-10}$~Pa\\
$p_\text{NE}^w$&toluene+$n$-hexane\footnote{Equimolar mixture, $T=298$~K, $\Delta{T}=25$~K}& $+2\times10^{-1}$~Pa &$+2\times10^{-3}$~Pa \\
$p_\text{NE}^w$& 1-methylnaphtalene+&+9~Pa&$+0.9\times10^{-1}$~Pa\\
        &n-heptane$^c$\\
$p_\text{NE}^w$&aniline + methanol$^c$&$-3\times10^{-1}$~Pa &$-3\times10^{-3}$~Pa \\
\botrule
\end{tabular*}
\label{T1}
\end{table}

In Table~\ref{T1} we present some order-of-magnitude estimates for the pressure $p_{{\rm{NE}}}^w$ induced by NE concentration fluctuations and compare the values with those for the original Casimir pressure $p_{{\rm{em}}}$, originating from electromagnetic fluctuations in vacuum~\cite{REF2}, and with the critical Casimir pressure $p_{{\rm{c}}}$~\cite{REF4,REF7}. The Casimir pressures induced by NE concentration fluctuations are much larger, than either $p_\text{em}$ or  $p_\text{c}$. Table~\ref{T1} also shows that $p_{{\rm{NE}}}^w$ can be either positive or negative, essentially depending on whether the concentration dependence of the heat of mixing is convex or concave.

Some authors have proposed NE Casimir forces induced by long-range fluctuations that originate from the spatial dependence of the noise correlations associated with the local fluctuation-dissipation theorem in the presence of a gradient~\cite{REF34,REF35}. However, for fluids it has been found that these NE correlations are insignificant compared to the NE correlations caused by a coupling between hydrodynamic modes in NE states considered here~\cite{REF36}.

Since, in contrast to the Casimir pressure $p_{{\rm{NE}}}^T$ induced by NE temperature fluctuations, the Casimir pressure $p_{{\rm{NE}}}^w$ induced by NE concentration exerts an actual force on the walls confining the liquid layer, we believe that the Casimir pressure induced by NE concentration fluctuations is a more promising candidate for an initial attempt to detect the phenomenon experimentally. As can be seen from Eq.~\eqref{E07}, the effect can be enhanced by selection of a mixture with a small diffusion coefficient $D$ and a large Soret coefficient $S_T$. This is the principal reason why in Table~\ref{T1} $p_{{\rm{NE}}}^w$ of 1-methylnaphtalene+n-heptane with $S_T  = 1.73 \times 10^{ - 2}~{\rm{K}}^{ - 1}$~\cite{REF37} is much larger than in toluene+n-hexane with $S_T  = 0.32 \times 10^{ - 2}~{\rm{K}}^{ - 1}$~\cite{REF38}. The validity of linear non-equilibrium fluctuating dynamics for the NE temperature and NE concentration fluctuations has been confirmed experimentally both by light scattering~\cite{REF12,REF12,REF39} and by shadowgraph experiments~\cite{REF40}.  Experimental evidence for the existence of a NE Casimir pressure would provide evidence for the validity of non-equilibrium nonlinear fluctuating hydrodynamics.

We note that similar NE concentration fluctuations and, hence NE Casimir forces, will also be present in liquid films with an isothermal concentration gradient or chemical-potential gradient~\cite{REF10,REF41,REF42,REF43}. Hence this kind of NE Casimir forces may be ubiquitous in nature.

The authors acknowledge valuable discussions with Jeremy N. Munday. The research at the University of Maryland was supported by the US National Science Foundation under Grant No. DMR-1401449.


\begin{thebibliography}{}
\bibitem{REF1}M. Kardar and R. Golestanian, Rev. Mod. Phys. $\bf{71}$, 1233 (1999).
\bibitem{REF2}H.B.G. Casimir, Proc. Kon. Ned. Akad. Wet. B $\bf{51}$, 793 (1948).
\bibitem{REF3}G.L. Klimchitskaya, U. Mohideen, and V.M. Mostepanenko, Rev. Mod. Phys. $\bf{81}$, 1827 (2009).
\bibitem{REF4}M.E. Fisher and P.-G. de Gennes, C.R. Acad. Sci. Ser. B $\bf{287}$, 207 (1978).
\bibitem{REF5}M. Krech, J. Phys. Condens. Matter $\bf{11}$, R391 (1999).
\bibitem{REF6}A. Gambassi, A. Maciolek, C. Hertlein, U. Nellen, L. Helden, C. Bechinger, and S. Dietrich, Phys. Rev. E $\bf{80}$, 061143 (2009).
\bibitem{REF7}A. Gambassi, C. Hertlein, L. Helden, S. Dietrich, and C. Bechinger, Europhys. News $\bf{40}$(1), 18 (2009).
\bibitem{REF8}J.R. Dorfman, T.R. Kirkpatrick, and J.V. Sengers, Ann. Rev. Phys. Chem. $\bf{45}$, 213 (1994).
\bibitem{REF9}R. Schmitmann and R.K.P. Zia, in \textit{Phase Transitions and Critical Phenomena}, Vol. 17, edited by C. Domb and J.L. Lebowitz (Academic, New York, 1995).
\bibitem{REF10}J.M. Ortiz de Z\'arate, and J.V. Sengers, \textit{Hydrodynamic Fluctuations in Fluids and Fluid Mixtures} (Elsevier, Amsterdam, 2006).
\bibitem{REF11}B. Derrida, J. Stat. Mech. P07023 (2007).
\bibitem{REF12}W.B. Li, P.N. Segr\`e, R.W. Gammon, and J.V. Sengers, Physica A $\bf{204}$, 399 (1994).
\bibitem{REF13}W.B. Li, K.J. Zhang, J.V. Sengers, R.W. Gammon, and J.M. Ortiz de Z\'arate, J. Chem. Phys. $\bf{112}$, 9139 (2000).
\bibitem{REF14}T.R. Kirkpatrick, E.G.D. Cohen, and J.R. Dorfman, Phys. Rev. A $\bf{26}$, 995 (1982).
\bibitem{REF33}M. {Tr\"ondle}, L. Hamau, and S. Dietrich, J. Chem. Phys. $\bf{129}$, 124716 (2008).
\bibitem{REF15}T.R. Kirkpatrick, J.M. Ortiz de Z\'arate, and J.V. Sengers, Phys. Rev. E $\bf{89}$, 022145 (2014).
\bibitem{REF16}K. Kawasaki, Ann. Phys. $\bf{61}$, 1 (1970).
\bibitem{REF17}D. Bedeaux and P. Mazur, Physica $\bf{73}$, 431 (1974).
\bibitem{REF18}D. Bedeaux and P. Mazur, Physica $\bf{75}$, 79 (1974).
\bibitem{REF19}M.H. Ernst, E.H. Hauge, and J.M.J. van Leeuwen, J. Stat. Phys. $\bf{15}$, 7 (1976).
\bibitem{REF20}M.H. Ernst, E.H. Hauge, and J.M.J. van Leeuwen, J. Stat. Phys. $\bf{15}$, 23 (1976).
\bibitem{REF21}D. Forster, D.R. Nelson, and M.J. Stephen, Phys. Rev. A $\bf{16}$, 732 (1977).
\bibitem{REF22}P. Kovtun and L.G. Yaffe, Phys. Rev. D $\bf{68}$, 025007 (2003).
\bibitem{REF23}S. Caron-Huot and O. Saremi, JHEP $\bf{11}$, 013 (2010).
\bibitem{REF24}W.W. Wood, J. Stat. Phys. $\bf{57}$, 675 (1989).
\bibitem{REF25}M.G. Velarde and R.S. Schechter, Phys. Fluids $\bf{15}$, 1707 (1972).
\bibitem{REF26}W.B. Li, \textit{Rayleigh Scattering from Liquids, Liquid Mixtures, and Polymer Solutions in Nonequilibrium Steady States}, Ph.D. Thesis (University of Maryland, College Park, MD, 1996).
\bibitem{REF27}B.M. Law and J.C. Nieuwoudt, Phys. Rev. A $\bf{40}$, 3880 (1989).
\bibitem{REF28}J.V. Sengers and J.M. Ortiz de Z\'arate, Revista Mexicana de F\'{\i}sica $\bf{48}$, Supl. 1, 14 (2002).
\bibitem{REF29}S. Chandrasekhar, \textit{Hydrodynamic and Hydromagnetic Stability}, (Oxford University Press, Dover Edition, 1981).
\bibitem{miarxiv}J.M. Ortiz de Z\'arate, T.R. Kirkpatrick, and J.V. Sengers, arXiv:1505.01355v1 (2015).
\bibitem{REF39}P.N. Segr\`e, R.W. Gammon, J.V. Sengers, and B.M. Law, Phys. Rev. A $\bf{45}$, 714 (1992).
\bibitem{REF30}T.R. Kirkpatrick, J.M. Ortiz de Z\'arate, and J.V. Sengers, Phys. Rev. Lett. $\bf{110}$, 235902 (2013).
\bibitem{REF31}J.L. Parker, Langmuir $\bf{8}$, 551 (1992).
\bibitem{REF32}G. Bressi, G. Carugno, R. Onofrio, and G. Ruoso, Phys. Rev. Lett. $\bf{88}$, 041804 (2002).
\bibitem{REF34}A. Najafi and R. Golestanian, Europhys. Lett. $\bf{68}$, 776 (2004).
\bibitem{REF35}A. Aminov, Y. Kafri, and M. Kardar, arXiv: 1501.01006v1 (2015).
\bibitem{REF36}J.M. Ortiz de Z\'arate and J.V. Sengers, J. Stat. Phys. $\bf{115}$, 1341 (2004).
\bibitem{REF37}S. Hartmann, G. Wittko, F. Schock, W. Grosz, F. Lindner, W. K\"ohler, and K.I. Morozov, J. Chem. Phys. $\bf{141}$, 134503 (2014).
\bibitem{REF38}K.J. Zhang, M.E. Briggs, R.W. Gammon, and J.V. Sengers, J. Chem. Phys. $\bf{104}$, 6881 (1996).
\bibitem{REF40}C.J. Takacs, A. Vailati, R. Cerbino, S. Mazzoni, M. Giglio, and D.S. Cannell, Phys. Rev. Lett. $\bf{106}$, 244502 (2011).
\bibitem{REF41}A. Vailati and M. Giglio, Nature $\bf{390}$, 262 (1997).
\bibitem{REF42}D. Brogioli, A. Vailati, and M. Giglio, J. Phys.: Condens. Matter $\bf{12}$, A39 (2000).
\bibitem{REF43}A. Donev, T.G. Fai, and E. Vanden-Eijnden, J. Stat. Mech. P04004 (2014).
\end{thebibliography}
\end{document}